\documentclass[aps,prd,showkeys,amsmath,amssymb]{revtex4-2}
\usepackage{graphicx}% Include figure files
\usepackage{dcolumn}% Align table columns on decimal point
\usepackage{bm}% bold math
\begin{document}
\title{Causality of brane universe via the general bulk-based formalisms with the non-zero Schwarzschild mass}
\author{Molin Liu$^{1}$}
\email{mlliu@xynu.edu.cn}
\author{Xi Zhou$^{1}$}
\author{Xiangsheng Tan$^{1}$}
%\author{Hongbao Zhang$^{2}$}
%\email{hongbaozhang@bnu.edu.cn}
\affiliation{$^{1}$College of Physics and Electronic Engineering,
Xinyang Normal University, Xinyang, 464000, P. R. China}
%\\$^{2}$Department of Physics, Beijing Normal University, Beijing 100875, P. R. China}
\begin{abstract}
In brane-world scenarios, electromagnetic waves (EMWs) are confined to the brane, while gravitational waves (GWs) can propagate through the bulk spacetime. This fundamental difference has been exploited in multiple cosmological studies to address some issues, such as the well-known horizon problem. This paper reinvestigates the problem using general bulk-based formalisms, with specific focus on how the non-zero Schwarzschild mass modifies geodesic motion. Our results demonstrate that the Schwarzschild mass significantly modifies the gravitational-to-photon horizon ratio. In the low-energy regime, our analysis constrains the anti-de Sitter curvature radius, i.e. $l H_0 \lesssim 10^{-29}$. Our finding agrees quantitatively with prior work. In the high-energy regime, the gravitational-to-photon horizon ratio $r_g/r_\gamma$ increases by thirty orders of magnitude, reaching $10^{33}$. Subject to the nucleosynthesis constraint $\sigma^{1/4} < 1 MeV$, the ratio becomes $10^{40}$. In this region, we observe the phenomenon of graviton bouncing by the brane, a behavior that has also been documented in prior literatures. Based on the observation of approximately $5\%$ dark radiation during the nucleosynthesis epoch, we further derive constraints on the relevant model parameters. Our results demonstrate that the non-zero Schwarzschild mass profoundly affects brane-world gravity. Crucially, some resulting effects may provide mechanisms to address persistent challenges in standard cosmology.
\end{abstract}
\keywords{Extra dimensions, Brane universe, Causality, Graviton propagation}
\maketitle
%\tableofcontents
\section{Introduction}
The progress in GW detection has provided researchers with powerful new tools to probe fundamental questions in theoretical physics. Remarkably, GWs observations have the potential to yield definitive constraints on the existence of large extra dimensions. In the events GW150914/GBM150914 \cite{ref:Abbott1,ref:Connaughton} and GW170817/GRB170817A \cite{ref:Abbott2,ref:Goldstein,ref:Savchenko}, we find that gravitational wave (GW) signals consistently arrive at Earth earlier than their electromagnetic (EMW) counterparts. If we assume that GWs and EMWs are emitted simultaneously from their source, this presents a fundamental question: why would GWs propagate faster than EMWs in vacuum? Several mechanisms have been proposed to explain this phenomenon \cite{ref:Ishihara,ref:Caldwell,ref:Abdalla1}. Within brane-world cosmology, one possible resolution suggests GWs may traverse extra-dimensional bulk spacetime while EMWs remain brane-confined, creating an effective shortcut for gravitational signals \cite{ref:Yu}. Recent studies indicate that multi-messenger gravitational wave astronomy could detect the distinctive interplay between the uniform ordinary dimensions and the warped geometry of the fifth dimension in brane-world scenarios \cite{ref:Chung2}.

One significant application of this gravitational wave shortcut mechanism is its potential to resolve the horizon problem in standard cosmology. Chung and Freese demonstrated that, in a lower-dimensional spacetime, causally disconnected regions could become connected through higher-dimensional null geodesics \cite{ref:Chung}. They further demonstrated that these higher-dimensional null geodesics could provide a resolution to the horizon problem in standard cosmology. However, a significant challenge for this approach is the lack of an identifiable, physically realistic spacetime configuration. Ishihara demonstrated that when the sum of density and pressure on the brane is positive, the resulting extrinsic curvature induces concave bending of the brane toward the bulk. Under these conditions, null geodesics of this type can naturally exist in the brane-world scenario \cite{ref:Ishihara}. Subsequently, Caldwell and Langlois investigated the separation between gravitational and photon horizons in 
the Anti-de Sitter-Schwarzschild (AdS-Sch) bulk, deliberately neglecting the effects of the Schwarzschild mass parameter $\mu$ \cite{ref:Caldwell}. Their findings suggest that the distinction between these two types of horizons is negligible, rendering it insufficient to address the horizon problem. For comprehensive reviews of recent developments in brane-world cosmology, see References \cite{ref:Maartens,ref:Langlois,ref:Brax}.

Recent studies have leveraged multi-messenger observations of GW events to investigate fundamental physics \cite{ref:Abbott3}, including the weak equivalence principle (WEP) \cite{ref:WEP}, Lorentz violation \cite{ref:LV}, and existence of large extra dimensions \cite{ref:LED}, along with other theoretical implications. Yu et al. revisited this shortcut mechanism by considering a spacetime with non-zero constant curvature $k$, while maintaining the Schwarzschild mass parameter $\mu$ at zero \cite{ref:Yu}. Their analysis demonstrates that the observed time delay between GWs and EMWs remains insufficient to resolve the horizon problem.

In this work, we focus on the case of the non-zero Schwarzschild mass parameter $\mu$ originating from the higher-dimensional bulk geometry. This mass parameter was first derived as an integration constant from the $0-0$ and $y-y$ components of the bulk Einstein field equations through application of the Israel junction conditions \cite{ref:Langlois,ref:Israel}. Subsequent work has established that this parameter couples to the bulk Weyl tensor and manifests effective mass characteristics, representing the five-dimensional generalization of the 
Schwarzschild mass. Critically, both the Caldwell-Langlois \cite{ref:Caldwell} and Yu et al. \cite{ref:Yu} analyses assume a vanishing Schwarzschild mass parameter is zero in their respective frameworks. As a fundamental parameter in brane-world cosmology, it is crucial to examine how $\mu$ affects the GW and EMW horizon radii within the brane universe. This fundamental question constitutes the principal motivation for the present investigation. To resolve this question, we initiate our analysis from the most general bulk-geometric framework. Employing a Lagrangian formulation, we derive the complete geodesic equations for test particles in a dynamically evolving brane with constant spatial curvature, embedded within a bulk of negatively curved bulk spacetime.

The structure of this paper is organized as follows: In Section \ref{Section2}, we establish the theoretical background and mathematical framework. In Section \ref{section3}, we employ the Lagrangian formalism to systematically derive the equations of motion for test particles in the brane-world scenario. In Section \ref{section4}, we compute the effective horizon radii for both gravitons and photons, analyzing the low-energy regime for gravitational waves and the high-energy limit for electromagnetic radiation. In Section \ref{section5}, we investigate the dynamics of brane trajectories and graviton propagation. Subsequently, in Section \ref{section6}, we examine the effects of dark radiation on horizon evolution. Finally, in Section \ref{section7}, we summarize our key findings, discuss their theoretical implications, and present concluding remarks.

\section{\label{Section2}The background}
We begin by considering an established brane-world scenario where a conventional Friedmann-Robertson-Walker (FRW) universe is embedded in the AdS-Sch bulk. The spacetime geometry can be characterized using the bulk-based metric formalisms \cite{ref:Ida,ref:Bowcock,ref:Maartens},
\begin{equation}\label{bulkmetric}
d S^2 = -f(R) d T^2 + \frac{1}{f(R)} d R^2 + R^2 d \Sigma_k^2,
\end{equation}
where the maximally symmetric subspace can be expressed as
\begin{equation}\label{Sigmak}
d \Sigma_k^2 = \frac{1}{1 - k r^2} d r^2 + r^2 d \theta^2 + r^2 \sin^2 \theta d \varphi^2.
\end{equation}
In this framework, the curvature radius of our three-dimensional spatial hypersurface $(r, \theta, \varphi)$ is governed by the parameter $k$ where $k = 0$ describes a flat space, $k = 1$ characterizes a spherical space, and $k = -1$ represents a hyperbolic space. The metric function $f(R)$ is given by
\begin{equation}\label{fRfunction}
f (R) = k + \frac{R^2}{l^2} - \frac{\mu}{R^2},
\end{equation}
where $\mu$ is the Schwarzschild mass induced by the extra dimension, $\Lambda$ is the bulk cosmological constant, and $l$ is the curvature radius of anti-de Sitter space, defined as $l \equiv \sqrt{-6/\Lambda}$.

Owing to the higher-dimensional spacetime geometry, the five-dimensional Schwarzschild mass parameter $\mu$ exhibits an $R^2$ dependence, in contrast to the conventional four-dimensional Schwarzschild mass's linear $R$ dependence. The metric in (\ref{bulkmetric}) describes the AdS-Sch bulk. A key advantage of this framework is its elegant embedding of the standard FRW cosmology into the brane-world scenario through a mathematically concise formulation. Notably, this solution remains universally valid regardless of $Z_2$ symmetry constraints about the brane. This coordinate system fundamentally differs from the conventional Gaussian normal (GN) formulation, as it adopts a bulk-centric perspective rather than being brane-constrained. Within brane-world cosmology, two principal theoretical frameworks have emerged. The first one is the brane-fixed approach, which treats the brane as a static boundary in a dynamic spacetime. The other approach is the bulk-fixed approach, where the brane follows a timelike trajectory through a stationary bulk geometry. This framework enables direct derivation of the brane's cosmological evolution equations, with the resulting solutions exactly describing the brane's dynamical trajectory through the AdS bulk spacetime. Meanwhile, in this brane-world scenario, the induced metric on the brane can be expressed as a function of proper time $t$,
\begin{equation}\label{propertime}
d t^2 = f(R_b) dT^2 - \frac{d R_b^2}{f(R_b)},
\end{equation}
where $f(R_b)$ is a fixed metric function defined in Eq.(\ref{fRfunction}), and $R_b$ denotes the brane's position. It can be identified with the standard cosmological scale factor $a(t)$ in the brane-world scenario. As a result, the induced metric on the brane takes the form of a FRW universe.
\section{\label{section3}The equation of motion}
\subsection{The Lagrangian of test particles}
We begin with the classical variational method for particle actions, focusing on two distinct particle types: gravitons and photons. In the brane-world scenario, photons are confined to the brane, whereas gravitons propagate through the bulk, including the fifth dimension. The equations of motion are obtained from the action principle
\begin{equation}\label{action}
S = \int \mathcal{L} d \xi,
\end{equation}
where $\xi$ is an affine parameter and $\mathcal{L}$ denotes the particle Lagrangian
\begin{equation}\label{Lagrangian}
2 \mathcal{L} \equiv -f(R) \dot{T}^2 + \frac{1}{f(R)} \dot{R}^2 + \frac{R^2}{1 - k r^2} \dot{r}^2 + r^2 R^2 \dot{\theta}^2 + r^2 \sin^2\theta R^2 \dot{\varphi}^2.
\end{equation}
Here, the overdot denotes differentiation with respect to the affine parameter $\xi$ along the geodesics. The Lagrangian satisfies the Euler-Lagrange equation
\begin{equation}\label{Euler-Lagrangeequation}
\frac{\partial \mathcal{L}}{\partial x^\mu} - \frac{d P_\mu}{d \xi} = 0.
\end{equation}
We adopt a spherical coordinate system $(r, \theta, \varphi)$ anchored on the brane, with the signal source $A$ positioned at the coordinate origin.EMWs and GWs propagate through different dimensions—EMWs confined to the brane and GWs traversing the bulk—before arriving at the receiver $B$ on Earth. Crucially, following Ishihara’s information propagation framework \cite{ref:Ishihara}, the receiver will first detect the GW signal at point $R$ followed by the EMW signal at point $Q$. Consequently, $R$ is in the past relative to $Q$. For a more detailed explanation, please refer to Ref.\cite{ref:Ishihara}.

To simplify the analysis, we may neglect the angular coordinates $(\theta, \varphi)$ by setting $d \theta = d \phi = 0$ in the Lagrangian (\ref{Lagrangian}). Consequently, the metric (\ref{bulkmetric}) reduces to the form
\begin{equation}\label{bulkmetric-new}
d S^2 = -f(R) d T^2 + \frac{1}{f(R)} d R^2 + \frac{R^2}{1 - k r^2} d r^2.
\end{equation}
Therefore, this approach enables us to express the test particle momenta as
\begin{eqnarray}
\label{PT}  P_T &=& \frac{\partial\mathcal{L}}{\partial(d T/d\xi)} = -f(R) \frac{d T}{d\xi},\\
\label{PR}  P_R &=& \frac{\partial\mathcal{L}}{\partial(d R/d\xi)} = \frac{1}{f(R)} \frac{d R}{d\xi}, \\
\label{Pr}  P_r &=& \frac{\partial\mathcal{L}}{\partial(d r/d\xi)} = \frac{R^2}{1 - k r^2} \frac{d r}{d\xi}.
\end{eqnarray}
In this work, we focus specifically on the case where $k = 0$ and $\mu \neq 0$, corresponding to a Lagrangian formulation that excludes dependence on both the bulk time $T$ and the radial coordinate $r$. Consequently, the cyclic coordinates $T$ and $r$ each correspond to a Killing vector field, yielding two conserved momenta,
\begin{eqnarray}
\label{3-dtdxi}\frac{d T}{d\xi} &=& \frac{E l^2 R^2}{R^4 - l^2 \mu}, \\
\label{3ParaPR}\frac{d r}{d\xi} &=& \frac{P}{R^2}.
\end{eqnarray}
Furthermore, applying the null condition $g_{AB} \frac{d x^A}{d \xi}\frac{d x^B}{d \xi} = 0$ to the geodesic equations yields the tangent vector components along the radial direction
\begin{equation}\label{3dRdlambda}
\left(\frac{d R}{d \xi}\right)^2 = \left(\frac{\mu}{R^4} - \frac{1}{l^2}\right) P^2+ E^2.
\end{equation}
Therefore, we establish that the three equations,specifically Eqs. (\ref{3-dtdxi}), (\ref{3ParaPR}) and (\ref{3dRdlambda}), govern the test particle motion along the $T$, $r$ and $R$ directions respectively.
\subsection{The null geodesics}
Focusing on radial geodesics in the bulk spacetime, we derive from Eqs.(\ref{3ParaPR}) and (\ref{3dRdlambda})
\begin{equation}\label{4newdRdr}
\frac{d R}{d r} = \left[\mu + \frac{\alpha^2 R^4}{l^2 (1 - \alpha^2)}\right]^{\frac{1}{2}},
\end{equation}
where $\alpha = \sqrt{1- P^2/E^2 l^2}$ is constrained to the interval ($0 < \alpha < 1$). The general solution to Eq. (\ref{4newdRdr}) takes the form
\begin{equation}\label{inte-dzdr}
\left(\mathcal{R} - \mathcal{R}^2 \right)^{\frac{1}{4}}{}_2F_1 \left(\frac{1}{2}, 1, \frac{5}{4}, \mathcal{R} \right) \bigg |_{\mathcal{R} _A}^{\mathcal{R}_B} = - \gamma^{\frac{1}{4}} \mu^{\frac{1}{2}} r,
\end{equation}
where we define the dimensionless coordinate $\mathcal{R} = 1/(1 + \gamma R^4)$, with $\gamma = \alpha^2/l^2 (1 - \alpha^2) \mu$ being a characteristic scale parameter. In this framework, the dimensionless coordinate $\mathcal{R}$ is bounded in the range of $(0,1)$. The limit $\mathcal{R}\rightarrow 0$ corresponds to the late-time universe, while $\mathcal{R}\rightarrow 1$ represents the early cosmological epoch. The hypergeometric function $_2F_1 (1/2, 1, 5/4, \mathcal{R})$ appearing in the solution admits the series expansion
\begin{equation}\label{4ser-expan}
_2F_1 \left(\frac{1}{2}, 1, \frac{5}{4}, \mathcal{R} \right) = \sum_{n = 0}^{\infty} \frac{\left(\frac{1}{2}\right)_n (1)_n}{\left(\frac{5}{4}\right)_n} \frac{\mathcal{R}^n}{n!},
\end{equation}
where $(a)_n$ represents the Pochhammer symbol, defined as
\begin{equation}\label{4Poch-Sym}
(a)_n = a (a + 1)\cdots(a + n - 1) = \frac{\Gamma(a + n)}{\Gamma (a)}.
\end{equation}

We now analyze the geodesic equation in the $T$ direction. From Eqs. (\ref{3-dtdxi}) and (\ref{3dRdlambda}), we obtain
\begin{equation}\label{5dRdT}
\frac{d R}{d T} = \frac{\alpha(R^4 - l^2 \mu)}{l^2 R^2} \sqrt{1 + \frac{1}{\gamma R^4}}.
\end{equation}
Integration of Eq. (\ref{5dRdT}) yields the solution form
\begin{equation}\label{5zTBA}
\mathcal{R}^{\frac{1}{4}} F_1 \left(\frac{1}{4}, -\frac{1}{4}, 1, \frac{5}{4}, \mathcal{R}, (1 + l^2 \gamma \mu) \mathcal{R} \right)\bigg|_{\mathcal{R}_A}^{\mathcal{R}_B} = - \alpha \gamma^{-\frac{1}{4}} l^{-2} T_{AB},
\end{equation}
where $T_{AB} = T_B - T_A$ denotes the bulk proper time interval, and $F_1 \left(\frac{1}{4}, -\frac{1}{4}, 1, \frac{5}{4}, \mathcal{R}, (1 + l^2 \gamma \mu) \mathcal{R} \right)$ represents the Appell hypergeometric function
\begin{equation}\label{5AppellF11}
F_1 (\frac{1}{4}, -\frac{1}{4}, 1, \frac{5}{4}, \mathcal{R}, (1 + l^2 \gamma \mu) \mathcal{R}) = \sum_{m = 0}^{\infty}\sum_{n = 0}^{\infty} \frac{(\frac{1}{4})_{m + n}(-\frac{1}{4})_{m}(1)_{n}}{m!n!(\frac{5}{4})_{m+n}} \mathcal{R}^{(m + n)} \left(\frac{E^2 l^2}{P^2}\right)^n,
\end{equation}
where $(\frac{1}{4})_{m + n}$, $(-\frac{1}{4})_{m}$, $(1)_{n}$, $(\frac{5}{4})_{m+n}$ denote Pochhammer symbols, whose properties are determined by the recurrence relation given in Eq. (\ref{4Poch-Sym}).
\section{\label{section4}The gravitational horizon and the photon horizon}
Given the analytical complexity of hypergeometric functions, we employ approximation methods to render the calculations more tractable while maintaining physical fidelity. In our analysis of late-universe dominated particle geodesics, we specifically investigate the asymptotic limits $R \rightarrow \infty$ and $\mathcal{R} \rightarrow 0$, which represent physically distinct regimes.Consequently, the hypergeometric functions appearing in Eqs.(\ref{4ser-expan}) and (\ref{5AppellF11}) admit series expansions about the point $\mathcal{R} = 0$, enabling analytical treatment of the local singularity regime,
\begin{equation}\label{3rRgeoEQ3}
\left(\frac{1}{R_A} - \frac{1}{R}\right) - \left(\frac{1}{R_A^5} - \frac{1}{R^5}\right) \frac{\mu P^2}{10 \alpha^2 E^2} = \frac{\alpha E}{P} r = \frac{\alpha}{l^2} \left(T - T_A\right).
\end{equation}

This analysis reveals a fundamental dispersion relation between the energy $E$ and momentum $P$, given by $P/E = r l^2/(T - T_A)$. In contrast to the case with $\mu = 0$, $k = 0$, we find the energy-momentum ratio $P/E$ becomes constant throughout the evolution. This demonstrates that the Schwarzschild mass $\mu$ does not affect the $P/E$ ratio, implying this energy-momentum relationship may hold universally for all brane-localized particles. This behavior originates fundamentally from the cyclic nature of both $T$ and $r$ coordinates in our framework.

Using Eq. (\ref{3rRgeoEQ3}), we eliminate the parameters $E$ and $P$ to obtain
\begin{equation}\label{3TRrgeoseed2}
 \left(\frac{1}{R_A} - \frac{1}{R}\right)^2 \left[\left(T - T_A\right)^2 - r^2 l^2\right] - \frac{\mu}{5} r^2 l^4 \left(\frac{1}{R_A^5} - \frac{1}{R^5}\right) \left(\frac{1}{R_A} - \frac{1}{R}\right)
 = \left[\left(T - T_A\right)^2 - r^2 l^2\right]^2 \frac{1}{l^4}.
\end{equation}
This constitutes the fundamental equation underlying our analysis. Therefore, for a graviton propagating from source $A$ in the host galaxy to receiver $B$ on Earth, the gravitational horizon radius $r_g$ is given by
\begin{equation}\label{grav-hori-rad}
r_g = \left[\eta^2 l^{-2}- \zeta^2 l^2 - \frac{\mu}{5} \zeta \chi l^4\right]^{\frac{1}{2}}.
\end{equation}
The parameters ($\zeta$, $\eta$, $\chi$) are defined as follows: $\eta = T_B - T_A$, $\zeta = 1/a_A - 1/a_B$ and $\chi = 1/a^5_A - 1/a^5_B$ where $a_A = R_A$ and $a_B = R_B$ denote the standard cosmological scale factors at points $A$ and $B$, respectively.

We emphasize that Eq. (\ref{grav-hori-rad}) constitutes the central result of this work, encapsulating the fundamental relationship between energy and momentum in our brane-world scenario. This horizon structure governs the causal propagation of GWs between brane points via the bulk dimension, defining their allowed signal connectivity
\begin{equation}\label{lum-radius}
r_\gamma = \int_{t_A}^{t_B} \frac{d t}{a}.
\end{equation}
This trajectory corresponds to photon propagation constrained to the brane. In subsequent analysis, we systematically examine both the gravitational horizon radius $r_g$ and electromagnetic horizon radius $r_\gamma$ across two distinct regimes: the low-energy limit ($lH\ll1$) and the high-energy limit($lH\gg1$).
\subsection{The low energy regime}
The low-energy regime ($lH\ll1$) describes a standard FRW-dominated universe. The brane matter content follows a perfect fluid equation of state $\omega =P/\rho$, where $\omega = 1/3$ characterizes radiation domination and $\omega = 0$ corresponds to pressureless matter. All cases obey the ideal fluid continuity equation. Thus, the Hubble parameter evolves as $H = H_B a_{B}^{\frac{3}{2}(1 + \omega)}a^{-\frac{3}{2}(1 + \omega)}$. The proper time interval between events $A$ and $B$ is given by
\begin{equation}\label{new4Dtimediff}
\eta = l H_B^{-1} a_B^{-1} a_A^{-\frac{1}{2}(3\omega + 1)}\left[\frac{2}{3\omega + 1} + \frac{\mu l^2}{3\omega - 7}\left(\frac{a_B}{a_A}\right)^{-4}\right].
\end{equation}
Substituting the proper time interval from Eq.(\ref{new4Dtimediff}) into Eqs.(\ref{grav-hori-rad}) and (\ref{lum-radius}) yields the ratio of gravitational to electromagnetic horizon radii
\begin{equation}\label{IVC-rgrg1}
\frac{r_g}{r_\gamma} \approx 1 + \frac{1}{2} \left(l H_B\right)^2 \frac{1 + 3 \omega}{5 + 3\omega} \left(\frac{a_B}{a_A}\right)^{(5+3\omega)/2} \mathcal{F}_1 +  \frac{1}{2} \frac{\mu l^2}{a_B^4}\frac{3\omega + 1}{3\omega - 7}\mathcal{F}_2,
\end{equation}
where
\begin{eqnarray}
\label{IVC-F1}\mathcal{F}_1 &=& \frac{(1 + 3\omega)(5 + 3\omega)\left[(a_A/a_B) - 1\right]^2}{4 [1 - (a_A / a_B)^{(1 + 3\omega)/2}]^2}\left(\frac{a_A}{a_B}\right)^{(1 + 3\omega)/2} + \frac{1 - (a_A/a_B)^{(5 + 3\omega)/2}}{1 - (a_A/a_B)^{(1 + 3\omega)/2}}, \\
\label{IVC-F2}\mathcal{F}_2 &=&  \frac{1 - (a_A/a_B)^{(3\omega - 7)/2}}{1 - (a_A/a_B)^{(3\omega + 1)/2}}.
\end{eqnarray}
In the massless limit $\mu \rightarrow 0$, the final term in Eq.(\ref{IVC-rgrg1}) vanishes, reproducing the 
low-energy regime result of Caldwell and Langlois \cite{ref:Caldwell}.

In the asymptotic limit $a_B \gg a_A$, the functional behavior of $\mathcal{F}_2$ is critically determined by the equation-of-state parameter $\omega$. Our analysis reveals two distinct regimes. For $\omega > 7/3$, the function asymptotically approaches unity $\mathcal{F}_2 = 1$. For $-1/3 <\omega < 7/3$, we obtain a power-law dependence $\mathcal{F}_2 = -(a_B/a_A)^{(7 - 3\omega)/2}$. This piecewise solution allows us to systematically compute the gravitational-to-electromagnetic horizon ratio $r_g/r_\gamma$ for all physical values of $\omega$.
\begin{equation}\label{IVC-ggrgafin}
\frac{r_g}{r_\gamma} = \begin{cases}
1 + \frac{1}{2} (l H_B)^2 \frac{1 + 3 \omega}{5 + 3\omega} \left(\frac{a_B}{a_A}\right)^{(5+3\omega)/2} + \frac{\mu l^2}{2 a_B^4}\frac{3\omega + 1}{3\omega - 7}, & \text{if } \omega > \frac{7}{3}, \\
1 + \frac{1}{2} (l H_B)^2 \frac{1 + 3 \omega}{5 + 3\omega} \left(\frac{a_B}{a_A}\right)^{(5+3\omega)/2} + \frac{\mu l^2}{2 a_B^4}\frac{3\omega + 1}{7 - 3\omega}\left(\frac{a_B}{a_A}\right)^{(7 - 3\omega)/2}, & \text{if }  -\frac{1}{3} <\omega < \frac{7}{3}.
\end{cases}
\end{equation}

To refine our statement, we provide insights into how the Schwarzschild mass influences geodesics in the low-energy regime.

$\bullet$ Our result for the ratio $r_g/r_\gamma$ effectively explains the geodesic behavior of gravitons and photons in the low-energy regime, independent of the equation-of-state parameter $\omega$, within the ranges $-1/3 <\omega < 7/3$ or $\omega >7/3$. Additionally, we find that as $\mu \rightarrow 0$, our results converge to those of the Caldwell-Langlois model.

$\bullet$ Our results are associated with $\mu l^2$ and $(H_B l)^2$. The influence of the extra-dimensional scale on these results is significant, as the coupling between the extra dimension, the Hubble parameter, and the Schwarzschild mass directly determines the physical behavior.

$\bullet$ The asymptotic behavior of geodesics is governed by the equation-of-state parameter $\omega$. The introduction of the Schwarzschild mass $\mu$ induces a bifurcation in the parameter space of $\omega$, producing two distinct physical regimes. We observe that for $\omega > 7/3$, the function $\mathcal{F}_2$ becomes constant, and consequently, the mass components of the solutions become independent of redshift (where $a_B/a_A = 1 + z$). A second key observation reveals that for $-1/3 < \omega < 7/3$, the function $\mathcal{F}_2$ exhibits redshift dependence, consequently inducing redshift variation in the mass component as well. However, the original interval $\omega\in(-1/3,+\infty)$ reported in Ref.\cite{ref:Caldwell} is now partitioned into two distinct 
regimes: $\omega\in(-1/3,7/3)$ and $\omega\in(7/3,+\infty)$.

$\bullet$ Considering a signal received at the present epoch $t_B = t_0$ and adopting $\omega = 0$ corresponding to the matter-dominated era, we derive the following result
\begin{equation}\label{IVC-rgrgomeg0}
\frac{r_g}{r_\gamma} \approx  1 + \frac{1}{10} (l H_0)^2 \left(1 + z\right)^{5/2} + \frac{1}{14}\frac{\mu l^2}{a_B^4}\left(1 + z\right)^{7/2},
\end{equation}
where $H_0$ denotes the present-day Hubble parameter, having a value of $H_0 \sim 2 h \times 10^{-42} GeV$. The parameter $z$ corresponds to the redshift of the signal source, with the signal propagation assumed to occur entirely during the matter-dominated cosmological epoch. The scale factor ratio is given by $a_B/a_A = 1 + z$. Current constraints from extra-dimensional theories \cite{ref:Hoyle,ref:Long} require the curvature radius $l$ of five-dimensional anti-de Sitter spacetime to satisfy $\sim 1 mm$. This constrains the curvature radius to $l\lesssim 1 mm \sim 5 \times 10^{12} GeV^{-1}$. Consequently, we obtain the dimensionless upper bound $l H_0 \lesssim 10^{-29}$ for the product of the curvature radius l and Hubble parameterH0.Furthermore, our analysis indicates that the dimensionless quantities $\mu l^2$ and $(H_B l)^2$ are of comparable magnitude, with both representing first-order perturbations. We thus obtain the constraint $\mu l^2 \sim (H_B l)^2 \lesssim 10^{-58}$. This confirms that the final term in Eq. (\ref{IVC-rgrgomeg0}) constitutes a higher-order perturbation. Thus, while the time delay grows with redshift $z$, its magnitude remains cosmologically negligible at present due to the extremely small dimensionless parameters $\mu l^2$ and $(H_B l)^2$.
\subsection{The high energy regime}
We now turn to the high-energy regime where $lH \gg 1$, corresponding to the early universe.In this regime, the energy density satisfies $\rho \gtrsim \sigma \approx M_{Pl}^2 \approx M_{(5)}^6/M_{Pl}^2$, where $\rho$ represents the brane tension. The temporal difference $T - T_A$ can then be expressed as
\begin{equation}\label{VBtimediff1}
T_B - T_A \approx l \int_{t_A}^{t_B} \frac{d t}{a} \left(Hl + \frac{1}{2H l} + \frac{\mu H l^3}{a^4}\right).
\end{equation}

Using the time difference expressed in Equation (\ref{VBtimediff1}), we can derive the ratio $r_g/r_\gamma$
\begin{equation}\label{VBratio1}
\frac{r_g}{r_\gamma} \approx \left[\frac{(2 + 3\omega)^2}{(5 + 6\omega)} \frac{(a_B/a_A - 1)(1 - (a_A/a_B)^{5 + 6\omega})}{(1 - (a_A/a_B)^{3\omega + 2})^2} + \frac{\mu l^4 H_B^2 (3\omega + 2)^2}{5 a_B^4} \frac{(a_B/a_A - 1)(a_B^5/a_A^5 - 1)}{(1 - (a_A/a_B)^{3\omega+2})^2} \right]^{\frac{1}{2}}.
\end{equation}

Clearly, in the limit where the mass vanishes $\mu \rightarrow 0$, the ratio reduces to the Caldwell-Langlois case \cite{ref:Caldwell}. The modified cosmological expansion law in high-energy brane-world scenarios necessitates a non-standard Friedmann equation \cite{ref:Binetruy1,ref:Binetruy2,ref:Langlois}
\cite{ref:Binetruy1,ref:Binetruy2,ref:Langlois}
\begin{equation}\label{VBcosexpan}
H^2 = \frac{\kappa_{(5)^4}}{36} \rho_{brane}^2 + \frac{\Lambda}{6}.
\end{equation}
In this framework, the five-dimensional gravitational constant $\kappa_{(5)}$ and mass scale $M_{(5)}$ are defined through the relations $\kappa_{(5)}^2 = M_{(5)}^{-3}$, while the four-dimensional Planck mass satisfies $M_{Pl}^2 = M_{(5)}^{3} l$. For $\omega > -2/3$ in the limit $a_B \gg a_A$, the ratio in Eq.(\ref{VBratio1}) admits the reformulation
\begin{equation}\label{VBratio3}
\frac{r_g}{r_\gamma} \sim \frac{2 + 3\omega}{\sqrt{5 + 6\omega}} \sqrt{\frac{a_B}{a_A}} \left[1 + \frac{1}{10} (5 + 6\omega)\frac{\mu l^4 H_B^2}{a_B^4}\left(\frac{a_B}{a_A}\right)^5\right].
\end{equation}
This constitutes our principal result in the high-energy regime. We now highlight several distinctive characteristics of the ratio $r_g/r_\gamma$ in this regime.

$\bullet$ In the high-energy regime, both graviton and photon geodesics are well-defined and exhibit distinct characteristics. Our results provide particularly robust resolution of the Caldwell-Langlois scenario. In the limit where the Schwarzschild mass vanishes ($\mu \rightarrow 0$), the ratio $r_g/r_\gamma$ (\ref{VBratio3}) exactly reproduces the Caldwell-Langlois solution.

$\bullet$ The second parenthetical term in Eq. (\ref{VBratio3}) explicitly contains the anti-de Sitter curvature radius $l$, revealing the direct influence of the bulk geometry on the brane dynamics. This demonstrates that the ratio $r_g/r_\gamma$ exhibits strong $l$-dependence in the high-energy regime, highlighting the crucial role of the AdS curvature scale in brane-world gravity. These results demonstrate that the extra dimension significantly modifies horizon dynamics, with particularly pronounced effects in the early universe's high-energy regime. This result stands in marked contrast to the Caldwell-Langlois scenario, where the ratio $r_g/r_\gamma$ maintains $l$-independence for $\mu = 0$ and $k = 0$, even at high energies.

$\bullet$ To simplify our analysis, we adopt the small-mass approximation by treating both 
$\mu l^2/a_B^4$ and $1/H_B^2 l^2$ as perturbational small quantities. This approximation guarantees 
the finiteness of the dimensionless ratio $(\mu l^2/a_B^4)/(1/H_B^2 l^2)$ in the high-energy limit. For computational simplicity, we normalize this limit to unity. Consequently, Eq. (\ref{VBratio3}) admits the following simplified form
\begin{equation}
\label{VBratio4} \frac{r_g}{r_\gamma} \sim \frac{2 + 3\omega}{\sqrt{5 + 6\omega}} \sqrt{\frac{a_B}{a_A}} \left[1 + \frac{5 + 6 \omega}{10}\left(\frac{a_B}{a_A}\right)^5\right].
\end{equation}

Our analysis reveals that as $a_A \rightarrow 0$ (characteristic of the early universe), the ratio $r_g/r_\gamma$ diverges ($r_g/r_\gamma \rightarrow \infty$). Consequently, this implies there is a lower bound on the duration for which the physics of this scenario remains valid. In standard cosmology, this limiting time corresponds to the Planck time \cite{ref:Caldwell}. In the model with extra dimensions, however, the limiting time is associated with the fundamental mass scale of the theory, $M_{(5)}$. Here, we employ a strategy similar to that used in the Caldwell-Langlois case, utilizing the limit to establish the applicability of this result. In this manner, the maximum ratio $r_g/r_\gamma$ is achieved when $t_B \sim l$ and $t_A \sim M_{(5)}$ where we have $M_{(5)} = (M_{pl}^2/l)^{1/3} \sim 3 \times 10^8 GeV$, $l \sim 1 mm = 5 \times 10^{12} GeV^{-1}$ and $M_{pl} = 1.2 \times 10^{19} GeV/c^2$. For radiation, the equation of state is given by $\omega = 1/3$ and the Hubble parameter behaves as $H \sim a^{-4}$. Therefore, with the minimum value of$M_{(5)} \sim 10^8 GeV$, we obtain a maximum ratio of $a_B/a_A \sim 10^6$. Thus, if we apply the time limit from the Caldwell-Langlois case, the mass will yield a new result: $r_g/r_\gamma \sim 10^{33}$.

In standard cosmology, the horizon problem emerges from the universe's gradual expansion distant regions that appear causally connected today were never in thermal equilibrium, as their particle horizons remained disjoint in the early universe \cite{ref:Gorbunov}. We can therefore compute the ratio of cosmological horizon radii between the present epoch $t_0$ and an earlier time $t_B$
\begin{equation}
\label{VBratio4ad} \frac{r_{\gamma 0}}{r_{\gamma B}} = \frac{\int_{0}^{t_0}\frac{dt}{a(t)}}{\int_{0}^{t_B}\frac{dt}{a(t)}} =\frac{\int_{0}^{a_0} d a/\left[a^2 H(a)\right]}{\int_{0}^{a_B} d a/\left[a^2 H(a)\right]}\simeq \frac{a(t_B) H(t_B)}{a(t_0) H(t_0)}.
\end{equation}
Following the Caldwell-Langlois formalism \cite{ref:Caldwell} and considering the early-time limit $t_B \sim l$, we can find
\begin{equation}\label{VBratio5}
\frac{r_{\gamma 0}}{r_{\gamma B}} \approx \left[(l H_0)^2 (1 + z_{eq})\right]^{-1/4}.
\end{equation}

Clearly, this result is independent of the parameter $\mu$. It is important to note that $z_{eq}$ represents the redshift at which the densities of matter and radiation are equal. It is at least several times greater than the redshift at the time of photon decoupling from matter. As a conservative estimate, we can assume the redshift to be $z_{eq} \sim 10^3$. For $l \sim 1$ mm, the ratio is given by $r_{\gamma 0}/r_{\gamma B} \sim 10^{14}$. Interestingly, the presence of mass parameters has significantly altered the results regarding causality, such that $r_g/r_\gamma \sim 10^{33} > r_{\gamma 0}/r_{\gamma B} \sim 10^{14}$. Therefore, we have a size difference of 19 orders of magnitude, which helps to address the classic horizon problem. In comparison to the previous result of $r_g/r_\gamma \sim 10^3$ reported in \cite{ref:Caldwell}, our calculations indicate that the bulk gravitational horizons could be increased by 30 orders of magnitude due to the presence of a non-zero 5D Schwarzschild mass.

We can then relax the constraints on $l$ derived from gravitational experiments. If we take into account an effective bulk cosmological constant that varies with time, $l$ could be reduced to millimeter scales following the nucleosynthesis constraints. Under this constraint, we have the condition $\sigma^{1/4} < 1 MeV$, which corresponds to the energy range relevant for nucleosynthesis. This implies that the upper limit of the fundamental mass scale $M_{(5)} \sim 10^4 GeV$. Thus, according to Eq.(\ref{VBratio4}), the ratio $r_g/r_\gamma$ could be raised to $10^{40}$. Meanwhile, with a reduced fundamental mass scale, the curvature radius of anti-de Sitter space will increase significantly, reaching approximately $l \sim 6 \times 10^{24} GeV^{-1}$. This suggests that the ratio $r_{\gamma 0}/r_{\gamma B}$ will decrease to $10^8$, as indicated by Eq.(\ref{VBratio5}). Therefore, we have $r_g/r_\gamma \sim 10^{40} > r_{\gamma 0}/r_{\gamma B} \sim 10^{8}$. Consequently, the radius of the gravitational horizon will increase by an additional 7 orders of magnitude under the nucleosynthesis constraint. Thus, we can conclude that the horizon problem could be further alleviated.

\section{\label{section5}Brane and Graviton trajectories in the bulk} 
Graviton trajectories in the high-energy ($\rho^2$) regime exhibit complex behavior due to both the brane's relativistic motion and multiple graviton-brane interactions, as demonstrated in Refs.\cite{ref:Langlois2,ref:Hebecker}). This section presents a comprehensive analysis of graviton trajectories in our framework.We compute the dynamical trajectories of the probe brane in the AdS-Schwarzschild background and analyze the corresponding graviton propagation in the bulk spacetime.

\subsection{\label{section5-1}The trajectories of the brane in AdS-Sch bulk}
By using the defined proper time $t$ (\ref{propertime}), which coincides with cosmic time on the brane, the brane trajectories are fully characterized by the coordinate functions $T(t)$ and $R(t)$. The brane velocity is given by $u^A = (\frac{d T(t)}{d t}, \frac{d R(t)}{d t}, 0)$. Applying the normalization condition $g_{AB} u^A u^B = -1$ yields
\begin{equation}\label{5-1normalization}
\frac{d T(t)}{d t} = \frac{\sqrt{f(R) + (d R/d t)^2}}{f(R)}, 
\end{equation}
where the metric is given by Eq.(\ref{bulkmetric-new}) with spatially flat geometry $k = 0$. This yields the brane's fundamental trajectory equation,
\begin{equation}\label{5-1originaltrabrane}
\frac{d R_b}{d T} = \frac{d R/d t}{d T/ dt} = \frac{f(R) d R/d t}{\sqrt{f(R) + (d R/d t)^2}}.
\end{equation}
Substituting the metric function $f(R)$ from Eq.(\ref{fRfunction}) into the above expression, we 
obtain
\begin{equation}\label{5-1originaltrabrane11}
\frac{d R_b}{d T} = \frac{\left(k + \frac{R^2}{l^2} - \frac{\mu}{R^2}\right) \frac{\dot{R}}{R}}{\sqrt{\frac{k}{R^2} + \frac{1}{l^2} - \frac{\mu}{R^4}+\left(\frac{\dot{R}}{R}\right)^2}}.
\end{equation}
For the case of $k=0$ and $\mu = 0$, the brane trajectory takes the form,
\begin{equation}\label{5-1originaltrabrane22}
\frac{d R_b}{d T} = \frac{R^2/l^2 (\dot{R}/R)}{\sqrt{1/l^2 + (\dot{R}/R)^2}}.
\end{equation}
Applying the scaling transformation $R \rightarrow R/l$ and $H \rightarrow l H$ yields
\begin{equation}\label{5-1originaltrabrane33}
\frac{d R_b}{d T} = R^2 \frac{H}{1 + H^2},
\end{equation}
which corresponds precisely to the $k = 0$, $\mu = 0$ case in Ref.\cite{ref:Langlois2}. This agreement confirms the validity of our computational framework. To maintain full transparency in our parameter space analysis and enable direct physical interpretation, we preserve the original parameterization without introducing rescaling conventions.

To determine the explicit form of the brane's trajectory, we must analyze the evolution of the Hubble parameter $H$, which is dictated by the Friedmann equation in this framework. We first apply the junction condition
\begin{equation}\label{5-1junctioncondi}
\left[h_A^C \nabla_C n_B\right] = \kappa_{(5)}^2 \left(\tau_{AB} - \frac{1}{3} \tau h_{AB}\right),
\end{equation}
where the square bracket denotes the jump across the brane, defined as $\left[Q\right] = Q(0^+) - Q(0^-)$. The quantity represents the discontinuity jump of $Q$ across the brane. The induced brane metric $h_{AB}$, given by Eq.(\ref{propertime}), governs the brane's geometry, while $\tau_{AB}$ describes the energy-momentum tensor of ordinary matter confined to the brane. Hence, the distributional brane energy-momentum tensor can be expressed as
\begin{equation}\label{5-1energymomentum1}
S_{AB} = - \sigma h_{AB} + \tau_{AB},
\end{equation}
where $\sigma$ represents the brane tension defined as $\sigma = \frac{\sqrt{-6\Lambda}}{\kappa_{(5)}^2}$. Note that using the relations $\kappa_{(5)}^2 = M_{(5)}^{-3} = l/M_{pl}^3$ and $\Lambda = -6/l^2$, we obtain the brane tension
$\sigma = \sqrt{-6\Lambda}/\kappa_{(5)}^2 = 6 M_{pl}^2/l^2$. Thus, we obtain
\begin{equation}\label{5-1H2-11}
H^2 = \frac{2 \rho}{l^2 \sigma} + \frac{\rho^2}{l^2 \sigma^2}=\frac{\rho}{3 M_{pl}^2} + \frac{l^2 \rho^2}{36 M_{pl}^4}.
\end{equation}
During radiation domination, the energy density evolves as $\rho = \rho_i R^{-4}$, where $\rho_i$ represents the density at the fiducial initial time $t_i$. By definition, when $\rho = \rho_i$, the scale factor $R = 1$. Substituting the expressions for $H$ and $\rho$ into the trajectory equation (\ref{5-1originaltrabrane22}), we obtain the following brane trajectory,
\begin{equation}\label{5-1branetrajectory1}
\frac{d R_b}{d T} = R^2 \frac{\sqrt{2\rho_i/R^4 + (\rho_i/R^4)^2}}{1 + \rho_i/R^4} \left[1 - \frac{\mu l^2}{R^4}\right]. 
\end{equation}
For the high-energy regime where $\rho_i/R^4 \gg 1$, the brane trajectory equation can be expressed as:
\begin{equation}\label{5-1hi1}
\frac{d R_b}{d T} \approx R^2 - \frac{\mu l^2}{R^2}.
\end{equation}
This equation can be derived through direct integration, yielding
\begin{equation}\label{5-1hi2}
T_b (R) = \frac{1}{4\sqrt{l}\mu^{1/4}}\left[2\arctan{\frac{R}{\sqrt{l}\mu^{1/4}}} + \log{(-R + \sqrt{l}\mu^{1/4})} -\log {(R + \sqrt{l}\mu^{1/4})}\right] + \text{const}.
\end{equation}
For the low-energy regime $\rho_i/R^4 \ll 1$, the brane trajectory equation reduces to
\begin{equation}\label{5-1low1}
\frac{d R_b}{d T} \approx \sqrt{2\rho_i}\left(1 - \frac{\mu l^2}{R^4}\right).
\end{equation} 
Through analogous integration to the high-energy case, we derive the brane trajectory for the low-energy regime
\begin{equation}\label{5-1low2}
T_b (R) = \frac{1}{4\sqrt{2\rho_i}} \left[4 R + \sqrt{l}\mu^{1/4} \left(-2\arctan{\frac{R}{\sqrt{l}\mu^{1/4}}} + \log{(-R + \sqrt{l}\mu^{1/4})} -\log {(R + \sqrt{l}\mu^{1/4})}\right) \right] + \text{const}.
\end{equation}
\subsection{\label{section5-2}The trajectories of the graviton in AdS-Sch bulk}
We now analyze the trajectories of gravitons propagating through the bulk. For context, we note that these geodesics were previously examined in earlier chapters. However, our present analysis focuses specifically on brane trajectories, with particular emphasis on the relativistic regime. In this work, we systematically investigate graviton-brane interactions across both high and low-energy regimes.

First, we recast the graviton trajectory equation (\ref{VBtimediff1}) in the following form,
\begin{equation}\label{5-2gravitoneq}
\frac{d R_g}{d T} = \frac{R}{l^2 \left(1 + \frac{\mu l^2}{R^4}\right)}.
\end{equation}
By integrating Eq.(\ref{5-2gravitoneq}), we derive the graviton trajectory in the high-energy regime,
\begin{equation}\label{5-2hi11}
T_g(R) = -\frac{l^4 \mu}{4 R^4} + l^2 \log{R} + \text{const}.
\end{equation}

In the low-energy regime, we employ the small-mass approximation ($\gamma \rightarrow \infty$ or $\mu \rightarrow 0$) to simplify calculations. Under this approximation, the trajectory equation (\ref{5dRdT}) reduces to
\begin{equation}\label{5-2gravitoneq22}
\frac{d R_g}{d T} = \frac{\alpha \left(R^4 - l^2 \mu\right)}{l^2 R^2}.
\end{equation}
Applying the same integration method to Eq.(\ref{5-2gravitoneq22}), we derive the graviton trajectory for the low-energy regime 
\begin{equation}\label{5-2low11}
T_g(R) = \frac{l^{3/2}}{4 \alpha \mu^{1/4}} \left(2\arctan{\frac{R}{\sqrt{l}\mu^{1/4}}} + \log{(-R + \sqrt{l}\mu^{1/4})} -\log {(R + \sqrt{l}\mu^{1/4})}\right) + \text{const}.
\end{equation}

\begin{figure*}
\centering
\includegraphics[width=0.4\columnwidth]{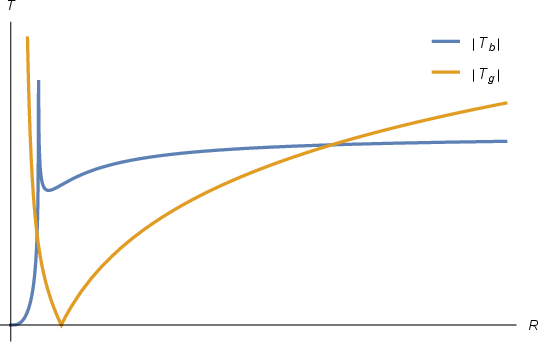}
\includegraphics[width=0.4\columnwidth]{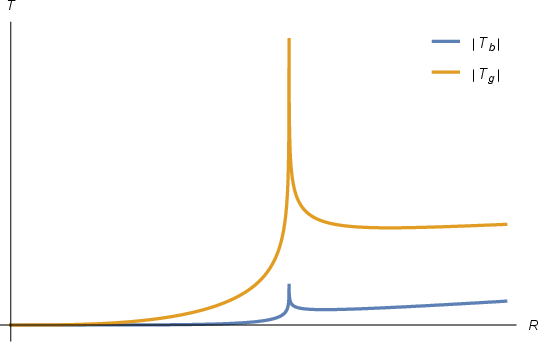}
\caption{\scriptsize Trajectories of the brane ($|T_b|$, blue curve) and graviton ($|T_g|$, yellow curve) in both high-energy (left panel) and low-energy (right panel) regimes. In the high-energy regime, gravitons emitted from the brane exhibit complex trajectories characterized by multiple intersections. During these interactions, gravitons tend to bounce at the brane interface. This behavior is consistent with established brane-world scenarios. In the low-energy regime, gravitons emitted from the brane exhibit minimal trajectory intersections.}
\label{fig1}
\end{figure*}

FIG.\ref{fig1} illustrates the brane and graviton trajectories, with the left panel showing the high-energy regime and the right panel displaying the low-energy regime. The plot reveals two distinct behaviors. In the low-energy regime, after graviton emission from the brane, their trajectories rarely intersect again. This suggests the absence of graviton bounce phenomena in this regime. In the high-energy regime, multiple brane-graviton trajectory intersections occur. Due to the brane's relativistic motion, gravitons inevitably bounce back at each 
intersection point.

\section{\label{section6}The influences of dark radiation on the horizons} 
As established in Refs.\cite{ref:Langlois2,ref:Hebecker,ref:Randall}, the bulk black hole in the RS2 model generates an effective dark radiation term in the four-dimensional Einstein equations. In this section, we examine the effects of dark radiation on both the gravitational and electromagnetic horizons, with special focus on its potential role in resolving the cosmological horizonproblem. Furthermore, we constrain the model parameters, such as parameter $\mu$, using observational signatures of dark radiation.

We begin our analysis with the Friedmann equation as derived in Ref.\cite{ref:Langlois2}, which provides the foundation for studying dark radiation effects,
\begin{equation}\label{6FriedmannEq}
H^2 = \left(\frac{\dot{R}}{R}\right)^2 = \left(\frac{\kappa_{(5)}^4}{36}\sigma^2 + \frac{\Lambda}{6}\right) + \frac{\kappa_{(5)}^4}{18} \sigma \rho + \frac{\kappa_{(5)}^4}{36} \rho^2 - \frac{k}{R^2} + \frac{\mu}{R^4},
\end{equation}
where dark radiation is characterized by the Weyl tensor contribution $\mu/R^4$. Incorporating the brane tension $\sigma$ as an additional energy density contribution ($\rho_{brane} = \sigma + \rho$), we derive the modified Friedmann equation for a flat three-space geometry ($k = 0$),
\begin{equation}\label{6FriedmannEq22}
H^2 = \frac{\kappa_{(5)}^4}{36}\rho_{brane}^2 - \frac{1}{l^2} + \frac{\mu}{R^4}.
\end{equation}
We note that our parameter conventions follow Ref.\cite{ref:Langlois}, which employs different notation from Ref.\cite{ref:Langlois2}. For instance, while the parameter $\mu$ in our work represents the Schwarzschild mass originating from the extra dimension, in Ref. \cite{ref:Langlois2} it instead characterizes a quantity related to the constant curvature radius $l$.

Next, we compute the dark radiation contribution to the Friedmann equation (\ref{6FriedmannEq22}). To quantitatively assess dark radiation's influence, we model the brane energy density $\rho_{brane}$ following conventional cosmology as $\rho_{brane} = C_0 R^{-3(1 + \omega)}$, where $\omega$ represents the equation of state parameter. For receiver $B$, we thus derive the following relation
\begin{equation}\label{6HB11}
H_B^2 = \frac{\kappa_{(5)}^4}{36} C_0^2 R_B^{-6(1+\omega)} - \frac{1}{l^2} + \frac{\mu}{R_B^4}.
\end{equation}
Through this derivation, we obtain the Friedmann equation in its modified form
\begin{equation}\label{6H222}
H^2 = \frac{1}{l^2} \left[H_B^2 l^2 - \frac{\mu l^2}{R_B^4} + 1\right] R_B^{6(1+\omega)}R^{-6(1+\omega)} - \frac{1}{l^2} + \frac{\mu}{R^4}.
\end{equation}
In the high-energy regime where $lH\gg1$ and $\mu l^2/R_B^4$ is a small quantity, the Friedmann equation (\ref{6H222}) simplifies to
\begin{equation}\label{6H233}
H^2 = H_B^2 R_B^{6(1+\omega)}R^{-6(1+\omega)} - \frac{1}{l^2} + \frac{\mu}{R^4}.
\end{equation}

Upon deriving the Friedmann equation that incorporates dark radiation, we can now examine the expressions for the gravitational horizon radius $r_g$ (\ref{grav-hori-rad}) and the electromagnetic horizon radius $r_\gamma$ (\ref{lum-radius}). A detailed analysis of these horizons has already been presented in earlier sections, so we will not repeat it here. Instead, we focus on investigating how dark radiation or equivalently, the modified Friedmann equation (\ref{6H233}) with dark radiation affects these two horizons. For the case where $\mu \neq 0$ and $\mu l^2/R^4 \ll 1$, we obtain the time difference $\eta$ given by
\begin{equation}\label{6eta11}
\eta = l \int_{tA}^t \frac{d t}{R} \sqrt{1 + H^2 l^2}.
\end{equation}
Meanwhile, we define $\xi = \int dR/R^2$ and $\chi = \int 5dR/R^6$. From these expressions, it is clear that dark radiation affects the time difference $\eta$ (\ref{6eta11}), but has no influence on $\xi$ and $\chi$. The gravitational horizon radius $r_g$ (\ref{grav-hori-rad}) can then be rewritten as
\begin{equation}\label{6rg11}
r_g = \left[\int\frac{d R}{R^2 H^2} \int\frac{d R}{R^2} - \frac{\mu}{5}\xi\chi l^4\right]^{1/2},
\end{equation}
and the electromagnetic horizon radius $r_\gamma$ from Eq. (\ref{lum-radius}) becomes
\begin{equation}\label{6rga11}
r_\gamma = \int \frac{d R}{R^2 H}.
\end{equation}
Equations (\ref{6rg11}) and (\ref{6rga11}) explicitly show that dark radiation modifies both the gravitational horizon radius $r_g$ and the electromagnetic horizon radius $r_\gamma$.

We now quantitatively assess the influence of dark radiation on these horizon radii. From Eq. (\ref{6FriedmannEq22}), where the parameter $\mu$ characterizes the dark radiation contribution, we observe that in the absence of dark radiation, the system reduces to
\begin{equation}\label{6H021212}
H_0^2 = \frac{\kappa_{(5)}^4}{36}\rho_{brane}^2 - \frac{1}{l^2}.
\end{equation}
Assuming the observed $5\%$ dark radiation contribution to $\rho^2$ reported in Ref.\cite{ref:Langlois2}, we derive $H^2 = H_0^2/95\%$, yielding $H = 1.03H_0$. Thus, when dark radiation is included, the gravitational horizon radius $r_g$ decreases, specifically scaling as $r_g \rightarrow 1/1.03 r_g$. Correspondingly, the electromagnetic horizon radius $r_\gamma$ also decreases, scaling as $r_{\gamma} \rightarrow 1/1.03 r_{\gamma}$. Given that both $r_{g}$ and $r_{\gamma}$ are decreasing concurrently, their relative ratio $r_{g}/r_{\gamma}$ remains unchanged. In other words, while dark radiation modifies both horizon radii proportionally, it is highly likely that it fails to resolve the cosmological horizon problem. Based on the observational constraints from Ref. \cite{ref:Hebecker}, which requires $\Omega_{d,N} \lesssim 0.05$ at nucleosynthesis, we determine that dark radiation production cannot exceed $5\%$ of the total energy density. Consequently, we derive the following constraint on the parameter $\mu$ at nucleosynthesis, i.e. $\mu/R_0^4 H_0^2 < 5\%$. Therefore, we obtain the constraint on the parameter $\mu < 5\% R_0^4/l^2 \sim 1.3 \times 10^{-51}GeV$ where $l \sim 6 \times 10^{24} GeV^{-1}$ at nucleosynthesis.
\section{\label{section7}Conclusion}
Finally, we provide a brief summary. With advancements in the study of GWs events, a significant amount of theoretical and experimental research has been conducted on various related GWs. It has become a trend for researchers to utilize GWs for in-depth investigations in theoretical physics.In current observations of GWs, it is often noted that GWs signals arrive at Earth earlier than those of EMWs. On the other hand, within brane world scenario, it is believed that ordinary matter is confined to the brane, while gravitons can propagate freely in the extra dimensions. It has been found that under certain conditions, such as when the sum of density and pressure on the brane is positive, or when the bulk possesses a negative cosmological constant, the extrinsic curvature may cause the brane to bend toward the bulk.In this manner, we can observe that the path taken by a graviton traveling from the bulk is shorter than that of a photon traveling from the brane. In this paper, we concentrate on how the Schwarzschild mass influences the geodesics of test particles in a bulk with a negative cosmological constant. We are particularly interested in determining whether this effect is significant enough to provide an explanation for the cosmological horizon problem.In this study, we employ the Lagrangian method to analyze the motion of test particles in brane world. Our findings indicate that if the Schwarzschild mass is non-zero in a bulk with a negative cosmological constant, even small mass values can significantly alleviate the cosmological horizon problem, particularly in the high-energy region corresponding to the early universe.It is important to note that there are two approaches to deriving cosmological solutions in brane world \cite{ref:Ida,ref:Kraus,ref:Bowcock}. One approach adopts a ``brane-based" perspective that utilizes GN coordinate system. In the brane-based approach, the Gauss-Codazzi formalism is employed to derive a pure brane Einstein equation, with the bulk geometry encoded in a single tensor $\mathcal{E}_{\mu\nu}$.The other approach adopts a ``bulk-based" perspective, providing the 5D AdS-Sch metric in which the cosmological solution is represented as a brane moving within AdS spacetime.In this paper, we investigate the motion of test particles within brane world from a ``bulk-based" perspective, where the trajectory of the brane is determined by its position in the fifth dimension. Naturally, the applicability of our conclusions to the ``brane-based" framework presents an intriguing research topic in its own right. There are numerous opportunities for interesting explorations in this area.

In this work, we investigate the dynamical scenario where both the brane and bulk gravitons undergo concurrent motion. The brane's trajectory is governed by both the bulk spacetime metric and Israel junction conditions. Our comparative analysis of low-energy and high-energy regimes (following Ref. \cite{ref:Langlois2}) reveals the substantial influence of the Schwarzschild mass $\mu$ on brane trajectory dynamics. As established in previous studies \cite{ref:Langlois2,ref:Hebecker}, bulk graviton propagation induces interactions with the brane tension. At relativistic brane velocities, graviton trajectories necessarily intersect the brane worldsheet multiple times, leading to repeated bounce interactions. Our calculations demonstrate that in the low-energy regime (right panel of FIG.\ref{fig1}), when the brane velocity becomes negligible compared to the graviton propagation speed following emission, subsequent interactions are kinematically forbidden, preventing any rebound phenomena. In contrast, within the high-energy regime (left panel of FIG.\ref{fig1}), relativistic brane motion leads to a high interaction probability, enabling recurrent graviton-brane collisions following initial emission. This generates repeated graviton-brane interactions, resulting in coherent back scattering of gravitons into the bulk spacetime. The process can iterate multiple times.

Within the brane-world paradigm, especially in AdS brane cosmology, the modified Friedmann equation on the brane differs significantly from standard cosmology. The modified Friedmann equation features a quadratic density term, see Eq.(\ref{6FriedmannEq}) along with a dark radiation component parameterized by the Weyl curvature. Our analysis reveals that dark radiation induces a proportional reduction in both horizon radii: the gravitational horizon $r_g$ (\ref{6rg11}) and electromagnetic horizon $r_\gamma$ (\ref{6rga11}). However, we find that dark radiation affects both horizons identically, leaving their relative ratio $r_g/r_\gamma$ invariant. Consequently, dark radiation cannot resolve the cosmological horizon problem, as it preserves the horizon ratio while proportionally reducing both scales.

\begin{acknowledgments}
We extend our gratitude to Hongbao Zhang and Enzhi Li for their invaluable discussions and kind helps. This research was partially funded by the National Natural Science Foundation (Grant No. 11475143).
\end{acknowledgments}


\begin{thebibliography}{*}
\bibitem{ref:Abbott1} B.P. Abbott, R. Abbott, T.D. Abbott, et al., Phys. Rev. Lett. 116 (2016) 061102, [arXiv:1602.03837[gr-qc]].
\bibitem{ref:Connaughton} V. Connaughton, E. Burns, A. Goldstein, et al., Astrophys. J. Lett. 826 (2016) L6, [arXiv:1602.03920[astro-ph.HE]].
\bibitem{ref:Abbott2} B.P. Abbott, R. Abbott, T.D. Abbott, et al., Phys. Rev. Lett. 119 (2017) 161101, [arXiv:1710.05832[gr-qc]].
\bibitem{ref:Goldstein} A. Goldstein, P. Veres, E. Burns, et al., ApJL 848 (2017) L14, [arXiv:1710.05446[astro-ph.HE]].
\bibitem{ref:Savchenko} V. Savchenko, C. Ferrigno, E. Kuulkers, et al., ApJL 848 (2017) L15, [arXiv:1710.05449[astro-ph.HE]].
\bibitem{ref:Ishihara}H. Ishihara, Phys. Rev. Lett. 86 (2001) 381-384, [arXiv:gr-qc/0007070].
\bibitem{ref:Caldwell} R. Caldwell and D.Langlois, Phys. Lett. B 511 (2001) 129, [arXiv:gr-qc/0103070].
\bibitem{ref:Abdalla1} E. Abdalla, B. Cuadros-Melgar, S.S Feng, B. Wang, Phys. Rev. D 65 (2002) 083512, [arXiv:hep-th/0109024].
\bibitem{ref:Yu} H. Yu, B.M. Gu, F.P. Huang et al, JCAP 1702 (2017) 039, [arXiv:1607.03388 [gr-qc]].
\bibitem{ref:Chung2} D.J.H. Chung, E.W. Kolb and A. Riotto, Phys. Rev. D 65 (2002) 083516, [arXiv:hep-ph/0008126].
\bibitem{ref:Chung} D.J.H. Chung and K. Freese, Phys. Rev. D 62 (2000) 063513, [arXiv:hep-ph/9910235].
\bibitem{ref:Langlois} D. Langlois, Prog. Theor. Phys. Suppl. 148 (2003) 181-212, [arXiv:hep-th/0209261].
\bibitem{ref:Israel} W. Israel, Nuovo Cim. B 44 (1966) 1,[Erratum-ibid. B 48 (1966) 463 ].
\bibitem{ref:Maartens} R. Maartens and K. Koyama, Living Rev. Relativity 13 (2010) 5, [arXiv:1004.3962 [hep-th]].
\bibitem{ref:Brax} P. Brax and C. Bruck, Class. Quant. Grav. 20 (2003) R201, [arXiv:hep-th/0303095].
\bibitem{ref:Abbott3} B.P. Abbott, R. Abbott, T.D. Abbott, et al., Phys. Rev. Lett. 116 (2016) 221101, [arXiv:1602.03841 [gr-qc]].
\bibitem{ref:WEP} C.M. Will, Living Rev. Relativ. 17 (2014) 4, [arXiv: 1403.7377 [gr-qc]]; J.J. Wei, H. Gao, X.F. Wu, P. Meszaros, Phys. Rev. Lett. 115 (2015) 261101, [arXiv:1512.07670 [astro-ph.HE]]; J.J. Wei, J.S. Wang, H. Gao, X.F. Wu, ApJL, 818 (2016) 1, L2, [arXiv:1601.04145 [astro-ph.HE]]; M. Liu, Z. Zhao, X. You et al., Phys. Lett. B 770 (2017) 8-15,[arXiv:1604.06668 [gr-qc]]; L. Yao, Z. Zhao, Y. Han et al., ApJ 900 (2020) 31, [arXiv:1909.04338[astro-ph.HE]].
\bibitem{ref:LV} S. Mirshekari, N. Yunes, C. M. Will, Phys. Rev. D 85 (2012) 024041, [arXiv:1110.2720 [gr-qc]]; Z. Wang, L. Shao, C. Liu, ApJ 921 (2021) 158,[arXiv:2108.02974 [gr-qc]].
\bibitem{ref:LED} N. Yunes, K. Yagi and F. Pretorius, Phys. Rev. D 94 (2016) 084002, [arXiv:1603.08955 [gr-qc]].
\bibitem{ref:Ida}D. Ida, JHEP 0009 (2000) 014, [arXiv:gr-qc/9912002].
\bibitem{ref:Bowcock} P. Bowcock, C. Charmousis, R. Gregory, Class.Quant.Grav. 17 (2000) 4745-4764,[arXiv:hep-th/0007177].
\bibitem{ref:Hoyle} C.D. Hoyle, et al., Phys. Rev. Lett. 86 (2001) 1418 [arXiv:hep-ph/0011014].
\bibitem{ref:Long} J.C. Long, H.W. Chan, A.B. Churnside, et al., Nature, 421 (2003) 922, [arXiv:hep-ph/0210004].
\bibitem{ref:Binetruy1}P. Binetruy, C. Deffayet and D. Langlois, Nucl. Phys. B 565 (2000) 269, [arXiv:hep-th/9905012].
\bibitem{ref:Binetruy2} P. Binetruy, C. Deffayet, U. Ellwanger, D. Langlois, Phys. Lett. B 477 (2000) 285, [arXiv:hep-th/9910219].
\bibitem{ref:Gorbunov} D. Gorbunov and V. Rubakov, World Scientific Publishing Co. Pte. Ltd (2011).
\bibitem{ref:Kraus} P. Kraus, JHEP 9912 (1999) 011 [arXiv:hep-th/9910149].
\bibitem{ref:Langlois2} D. Langlois and L. Sorbo, Phys. Rev. D 68 (2003) 084006, [arXiv:hep-th/0306281].
\bibitem{ref:Hebecker}A. Hebecker and J. March-Russell, Nucl. Phys. B 608 (2001) 375-393, [arXiv:hep-ph/0103214].
\bibitem{ref:Randall}L. Randall and R. Sundrum, Phys. Rev. Lett. 83 (1999) 4690, [arXiv:hep-th/9906064].
\end{thebibliography}
\end{document}